\documentclass[harvard,11pt]{article}
  \usepackage{amssymb}
\usepackage{verbatim}
\usepackage{multirow}
\usepackage{pifont}
\usepackage{graphicx}
\usepackage{tikz}
\usetikzlibrary{matrix}
\usepackage{lscape}
\usepackage{longtable}
\usepackage{listings}
\usepackage{amsfonts}
\usepackage{amsmath}
\usepackage{upgreek}
\usepackage{adjustbox}
\usepackage[bottom]{footmisc}

\usepackage{pstricks}
\usepackage{lipsum}
\usepackage{hyperref}
\hypersetup{
  colorlinks=true,
  linkcolor=blue,
  filecolor=black,      
  urlcolor=blue,
  citecolor=blue,	
}
\usepackage{placeins}
\usepackage{amsmath}
\usepackage[abs]{overpic}
\usepackage{linegoal}
\usepackage{bbm}
\usepackage{fontenc}
\usepackage{bm}
\usepackage{booktabs}
\usepackage[round]{natbib}
\usepackage{amsmath}
\usepackage{caption}
\usepackage{tabu}
\usepackage{tabularx,ragged2e}
\usepackage{amsmath}
\usepackage{algorithm}
\usepackage{physics}
\usepackage{booktabs}
\usepackage{algpseudocode}
\usepackage{graphicx,epstopdf}
\usepackage{setspace,caption}
\usepackage{rotating}
\usepackage{amsmath}
\usepackage[margin=1in]{geometry}
\usepackage{ragged2e}
\usepackage{listings}
\usepackage{subcaption}
\usepackage{xargs}
\usepackage[mathscr]{euscript}
\usetikzlibrary{decorations.pathreplacing,calligraphy}
\usepackage[hyphenbreaks]{breakurl}

\newcommandx{\unsure}[2][1=]{\todo[linecolor=red,backgroundcolor=red!25,bordercolor=red,#1]{#2}}
                                   \newcommandx{\change}[2][1=]{\todo[linecolor=blue,backgroundcolor=blue!25,bordercolor=blue,#1]{#2}}
                                                                      \newcommandx{\info}[2][1=]{\todo[linecolor=OliveGreen,backgroundcolor=OliveGreen!25,bordercolor=OliveGreen,#1]{#2}}
                                                                                                       \newcommandx{\improvement}[2][1=]{\todo[linecolor=Plum,backgroundcolor=Plum!25,bordercolor=Plum,#1]{#2}}
                                                                                                                                               \newcommandx{\thiswillnotshow}[2][1=]{\todo[disable,#1]{#2}}

                                                                                                                                                                                           \newcolumntype{L}{>{\RaggedRight}X}
                                                                                                                                                                                           \newcolumntype{R}{>{\RaggedLeft}X}
                                                                                                                                                                                           \newcommand*{\rom}[1]{\expandafter\@slowromancap\romannumeral #1@}
                                                                                                                                                                                             \newcommand{\R}{\mathbb{R}}

                                                                                                                                                                                               \renewcommand{\cite}{\citeasnoun}
                                                                                                                                                                                                 \numberwithin{equation}{section}
                                                                                                                                                                                                 
                                                                                                                                                                                                 \usetikzlibrary{decorations.pathreplacing,calc}

                                                                                                                                                                                                   \tikzset{brace/.style={decorate, decoration={brace}},
                                                                                                                                                                                                   brace mirrored/.style={decorate, decoration={brace,mirror}},
                                                                                                                                                                                                   }
                                                                                                                                                                                                   
                                                                                                                                                                                                   \newcounter{brace}
                                                                                                                                                                                                   \newcounter{arrow}
\usepackage{Sweave}
\begin{document}

	\title{{\bf Disaggregating Time-Series with Many Indicators: An Overview of the DisaggregateTS Package}}
	\author{
		Luke Mosley\thanks{STOR-i Centre for Doctoral Training, Department of Mathematics and Statistics, Lancaster University, United Kingdom}
		\and
		Kaveh Salehzadeh Nobari\thanks{Imperial College Business School, Imperial College London, United Kingdom}
	\and  
		Giuseppe Brandi\footnotemark[2]
        \and
        Alex Gibberd\footnotemark[1]
  }
	
	\date{\today
                                                                                                                                                                                                 }
                                                                                                                                                                                                 \maketitle

                                                                                                                                                                                                 \begin{abstract}
                                                                                                                                                                                                 Low-frequency time-series (e.g., quarterly data) are often treated as benchmarks for interpolating to higher frequencies, since they generally exhibit greater precision and accuracy in contrast to their high-frequency counterparts (e.g., monthly data) reported by governmental bodies. 
                                                                                                                                                                                                 %
                                                                                                                                                                                                 %
                                                                                                                                                                                                 An array of regression-based methods have been proposed in the literature which aim to estimate a target high-frequency series using higher frequency indicators. However, in the era of big data and with the prevalence of large volume of administrative data-sources there is a need to extend traditional methods to work in high-dimensional settings, i.e. where the number of indicators is similar or larger than the number of low-frequency samples. The package \textbf{DisaggregateTS} includes both classical regressions-based disaggregation methods alongside recent extensions to high-dimensional settings, c.f. \citet{10.1111/rssa.12952}. This paper provides guidance on how to implement these methods via the package in R, and demonstrates their use in an application to disaggregating CO2 emissions.
                                                                                                                                                                                                 \end{abstract}

                                                                                                                                                                                                 
                                                                                                                                                                                                 
                                                                                                                                                                                                 
                                                                                                                                                                                                 \section{Introduction\label{sec:Introduction}}
                                                                                                                                                                                                 
                                                                                                                                                                                                 Economic and administrative data, such as recorded surveys and consensus, are often disseminated by international governmental agencies at low or inconsistent frequencies, or irregularly-spaced intervals. To aid the forecasting of the evolution of the dynamics of these macroeconomic and socioeconomic indicators, as well as their comparison with higher resolution indicators provided by international agencies, statistical agencies rely on signal extraction, interpolation and temporal distribution adjustments of the low-frequency data to provide high precision and uninterrupted historical data. Although, temporal distribution, interpolation and benchmarking are closely associated with one another, this article and its respective package (\textbf{DisaggregateTS}), expend particular attention to interpolation and temporal distribution (disaggregation) techniques, where the latter is predicated on regression-based methods\footnote{See \citet{dagum2006benchmarking} for an overview of benchmarking, interpolation, temporal distribution and calendarization techniques.}. These regression-based temporal distribution techniques rely on high-frequency indicators to estimate (relatively) accurate high-frequency data points. With the prevalence of large volume of high-frequency administrative data, a great body of literature pertaining to statistical and machine learning methods has been dedicated to taking advantage of these additional resources for forecasting purposes (see \citet{fuleky2019macroeconomic} for an overview of macroeconomic forecasting in the presence of big data). Additionally, one may wish to utilize these abundant indicators to generate high-frequency estimates of low-frequency time-series with greater precision. However, in high-dimensional linear regression models where the number of dimensions surpass that of the observations, consistent estimates of the parameters is not possible without imposing additional structure (see \citet{wainwright2019high}). Hence, this article and the package \textbf{DisaggregateTS} adapt recent contributions in high-dimensional temporal disaggregation (see \citet{10.1111/rssa.12952}) to extend previous work within this domain (see the package \citet{sax2016package} and its corresponding article \citet{sax2013temporal2}) to high-dimensional settings.
                                                                                                                                                                                                 
                                                                                                                                                                                                 As noted by \citet{dagum2006benchmarking}, time-series data reported by most governmental and administrative agencies tend to be of low-frequency and precise, but not particularly timely, whereas their high-frequency counterparts seldom uphold the same degree of precision.
                                                                                                                                                                                                 The aim of temporal distribution techniques is to generate high-frequency estimates that can track shorter term movements, than directly observable with the direct low-frequency observations. While interpolation problems are generally encountered in the context of stock series, where say, the quarterly value of the low-frequency series must coincide with the value of third month of the high-frequency data (of the same quarter), temporal distribution problems often concern flow series, where instead the value of the low-frequency quarterly series must agree with the sum (or weighted combination) of the values of the high-frequency months in that quarter. The latter approach is generally accomplished by identifying and taking advantage of a number of high-frequency indicators which are deemed to behave in a similar manner to the low-frequency series, and by estimating the high-frequency series through a linear combination of such indicators. 
                                                                                                                                                                                                 
                                                                                                                                                                                                 In the last few decades, a significant number of articles have been published within this domain---see \citet{dagum2006benchmarking} for a detailed review of these techniques. Notable studies within this context include the additive regression-based benchmarking methods of \citet{denton1971adjustment} and \citet{chow1971best,chow1976best}, as well as those proposed by \citet{fernandez1981methodological} and \citet{litterman1983random} in the presence of highly persistent error processes. More recently these methods have been extended to the high-dimensional setting by \citet{10.1111/rssa.12952}, where prior information on the structure on the linear regression model is used to enable estimation, and better condition the regression problem. Specifically, this is accomplished by \textquotedblleft{}least absolute shrinkage and selection operator\textquotedblright{} (LASSO hereafter) proposed by \citet{tibshirani1996regression}, which in principle selects an appropriate model by penalizing the coefficients (in scale) of the high-dimensional regression, in effect discarding the irrelevant indicators from the model. In what follows, we demonstrate how to apply these methods using \textbf{DisaggregateTS} to easily estimate high-frequency series of interest.
                                                                                                                                                                                                 
                                                                                                                                                                                                 The remainder of the paper is organized as follows: Section \ref{sec:Sparse temporal disaggregation} introduces the methodologies underlining noteworthy temporal disaggregation techniques that have been included in the \textbf{DisaggregateTS} package, as well as their extensions to high-dimensional settings. Section \ref{sec:The DisaggregateTS package} introduces the \textbf{DisaggregateTS} package and some of its useful functions. Moreover, examples predicated both on simulations (using a function provided in the package that generates synthetic data) and empirical data are explored to familiarize the reader with the package and its functionality. Finally, the paper is concluded in Section \ref{sec:Summary}.
                                                                                                                                                                                                 
                                                                                                                                                                                                 \section{Sparse temporal disaggregation\label{sec:Sparse temporal disaggregation}}

                                                                                                                                                                                                 \subsection{Classical regression-based techniques\label{sec:Classical regression-based techniques}}

                                                                                                                                                                                                 \begin{figure}[hbtp!]
                                                                                                                                                                                                 \centering
                                                                                                                                                                                                 \includegraphics[width=\textwidth,keepaspectratio]{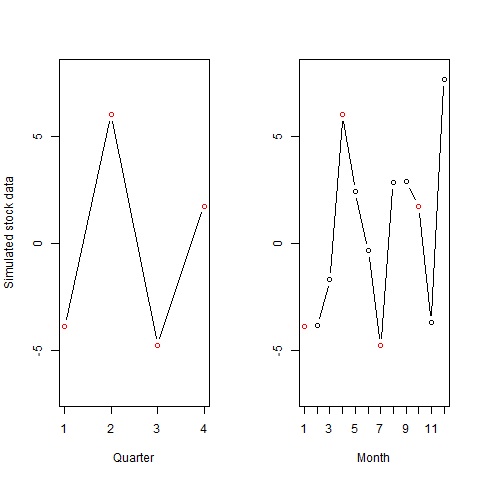}
                                                                                                                                                                                                 \caption{The data in these figures have been generated using the \texttt{TempDisaggDGP()} function of the \textbf{DisaggregateTS} package and represent simulated stock data. The value at each quarter of low-frequency data must coincide with that of the first month of that quarter in the corresponding high-frequency series, which are presented with red dots.}
                                                                                                                                                                                                 \end{figure}
                                                                                                                                                                                                 
                                                                                                                                                                                                 \begin{figure}[hbtp!]
                                                                                                                                                                                                 \centering
                                                                                                                                                                                                 \includegraphics[width=\textwidth,keepaspectratio]{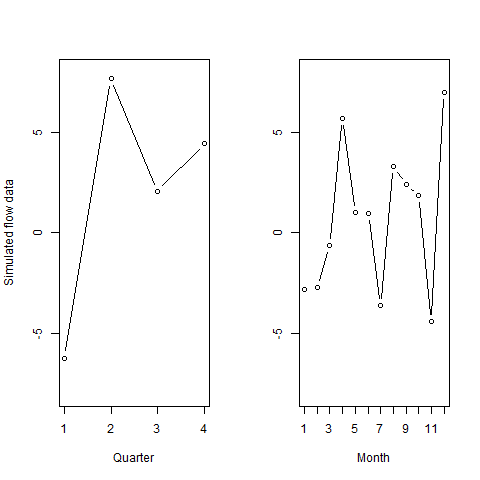}
                                                                                                                                                                                                 \caption{The data in these figures have also been generated using the \texttt{TempDisaggDGP()} function of the \textbf{DisaggregateTS} package and correspond to flow data.}
                                                                                                                                                                                                 \end{figure}

                                                                                                                                                                                                 
                                                                                                                                                                                                 Suppose we observe a low-frequency series, say, quarterly GDP, encoded as the vector $\bm{y}_q\in \mathbb{R}^n$, containing $n$ quarterly observations. We desire to disaggregate this series to higher frequencies (say monthly), where the disaggregated series is denoted $\bm{y}_m\in \R^p$, with $p=3n$. Furthermore, we wish that the disaggregated series be temporally consistent without exhibiting any jumps between quarters (see Section 3.4 of \citet{dagum2006benchmarking} for examples of such inconsistencies between the periods). The challenge is to identify an approach that distributes the variation between each observed quarterly point to the monthly level. A method that has been extensively studied in the literature concerns finding high-frequency (e.g., monthly) indicator series that are thought to exhibit similar inter-quarterly movements as the low-frequency variable of interest. Let us denote a set of $p$ observations from these $d$ indicators as the matrix $\bm{X}_m\in \mathbb{R}^{p\times d}$. A classical approach to provide high-frequency estimates is the regression-based temporal disaggregation technique proposed by \citet{chow1971best} whereby the unobserved monthly series $\bm{y}_m$ are assumed to follow the regression:
                                                                                                                                                                                                   \begin{equation}\label{eq: ChowLin}
                                                                                                                                                                                                 \bm{y}_m=\bm{X}_m\bm{\beta }+\bm{u}_m,\quad \bm{u}_m\sim N(\bm{0}_m,\bm{V}_m)
                                                                                                                                                                                                 \end{equation}
                                                                                                                                                                                                 where $\bm{\beta} \in\R^d$ is a vector of regression coefficients to be estimated (noting that $\bm{X}_m$ may contain deterministic terms) and $\bm{u}_m\in \R^p$ is a vector of residuals. \citet{chow1971best} assume that $\bm{u}_m$ follows as AR(1) process of the form $u_t=\rho u_{t-1}+\varepsilon_t$ with $\varepsilon_t\sim N(0,\sigma^2)$ and $\lvert \rho\rvert <1$. The assumption of stationary residuals allows for a cointegrating relationship between $\bm{y}_m$ and $\bm{X}_m$ when they are integrated of the same order. 
                                                                                                                                                                                                 Thus, the covariance matrix has a well-known Toeplitz structure as follows:
                                                                                                                                                                                                   \begin{equation}
                                                                                                                                                                                                 \bm{V}_m=\frac{\sigma^2}{1-\rho^2}
                                                                                                                                                                                                 \begin{pmatrix}
                                                                                                                                                                                                 1 & \rho &\cdots & \rho^{p-1}\\
                                                                                                                                                                                                 \rho & 1 &\cdots & \rho^{p-2}\\
                                                                                                                                                                                                 \vdots&\vdots&\ddots&\vdots \\
                                                                                                                                                                                                 \rho^{p-1} & \rho^{p-2} &\cdots & 1\\
                                                                                                                                                                                                 \end{pmatrix}
                                                                                                                                                                                                 \end{equation} 
                                                                                                                                                                                                 where $\rho$ and $\sigma$ are unknown parameters that need to be estimated. The dependent variable $\bm{y}_m$ in \eqref{eq: ChowLin} is unobserved, hence the regression is premultiplied by the $n\times p$ aggregation matrix $\bm{C}$, where:
                                                                                                                                                                                                   \begin{align}\label{eq: flowData}
                                                                                                                                                                                                 \begin{split}
                                                                                                                                                                                                 \bm{C}&=\bm{I}_n\otimes (1,1,1)\\
                                                                                                                                                                                                 &=
                                                                                                                                                                                                   \begin{pmatrix}
                                                                                                                                                                                                 1&1&1&0&0&0&0&\cdots &0\\
                                                                                                                                                                                                 0&0&0 & 1&1&1&0&\cdots &0\\
                                                                                                                                                                                                 \vdots&\vdots&\vdots & &\ddots&\ddots&\ddots&\vdots &\vdots\\
                                                                                                                                                                                                 0&\cdots&0&0&0&0&1&1 &1
                                                                                                                                                                                                 \end{pmatrix}_{n\times p}
                                                                                                                                                                                                 \end{split}
                                                                                                                                                                                                 \end{align}
                                                                                                                                                                                                 where the vector of ones in \eqref{eq: flowData} is used for flow data (e.g., GDP), such that the sum of the monthly GDPs coincides with its quarterly counterpart\footnote{For alternative aggregations see \citet{quilis2018temporal} and \citet{sax2016package}. For instance, if quarterly values correspond to averages of monthly values, then the vector in equation \eqref{eq: flowData} assumes the form $(0.33,0.33,0.33)$.}. The premultiplication yields the quarterly counterpart of \eqref{eq: ChowLin}:
                                                                                                                                                                                                   \begin{equation}
                                                                                                                                                                                                 \bm{C}\bm{y}_m=\bm{C}\bm{X}_m\bm{\beta }+\bm{C}\bm{u}_m,\quad \bm{C}\bm{u}_m\sim N(\bm{C}\bm{0}_m,\bm{C}\bm{V}_m\bm{C}^{\top}).
                                                                                                                                                                                                 \end{equation}
                                                                                                                                                                                                 The GLS estimator for $\bm{\beta}$ is thus expressed as follows:
                                                                                                                                                                                                   \begin{align}
                                                                                                                                                                                                 \hat{\bm{\beta}}=&\arg\min_{\bm{\beta}\in\R^d}\left\{\Big\lVert \bm{V}_q^{-\frac{1}{2}}(\bm{y}_q-\bm{X}_q\bm{\beta})\Big\rVert_2^2\right\}\label{eq:CostFunction}\\
                                                                                                                                                                                                 =&\left(\bm{X}_q^{\top}\bm{V}_q^{-1}\bm{X}_q\right)^{-1}\bm{X}_q^{\top}\bm{V}_q^{-1}\bm{y}_q\label{eq:GLSEstimator}
                                                                                                                                                                                                 \end{align}
                                                                                                                                                                                                 where $\bm{X}_q=\bm{CX}_m$, $\bm{y}_q=\bm{Cy}_m$ and $\bm{V}_q^{-1}=\bm{C}\bm{V}_m\bm{C}$. Note that estimating $\bm{\beta}$ requires the knowledge of the unknown parameters $\sigma$ and $\rho$ in $\bm{V}_m$ which are unknown. We employ the profile-likelihood maximization technique of \citet{bournay1979reflexions} which entails first estimating $\hat{\bm{\beta}}$ and $\bm{V}_q$ and subsequently conducting a grid-search over the range $\rho\in (-1, 1)$ for the autoregressive parameter. \citet{chow1976best} show the optimal solution is obtained by:
                                                                                                                                                                                                   
                                                                                                                                                                                                   \begin{equation}
                                                                                                                                                                                                 \hat{\bm{y}}_m=\bm{X}_m\hat{\bm{\beta}}+\bm{V}_m\bm{C}\bm{V}_q^{-1}\left(\bm{y}_q-\bm{X}_q\hat{\beta}\right),
                                                                                                                                                                                                 \end{equation}
                                                                                                                                                                                                 where $\bm{X}_m\hat{\bm{\beta}}$ is the conditional expectation of $\bm{y}_m$ given $\bm{X}_m$ and the estimate of the monthly residuals are obtained by disaggregating the quarterly residuals $\bm{y}_q-\bm{X}_q\hat{\beta}$ to attain temporal consistency between $\hat{\bm{y}}_m$ and $\bm{y}$.  
                                                                                                                                                                                                 
                                                                                                                                                                                                 Other variants of the regression-based techniques include those proposed by \citet{denton1971adjustment}, \citet{dagum2006benchmarking}, \citet{fernandez1981methodological} and \citet{litterman1983random}. The latter two consider scenarios where $\bm{y}_m$ and $\bm{X}_m$ are not cointegrated. Although, these traditional techniques are included in the \textbf{DisaggregateTS} package, for an overview of different temporal disaggregarion techniques and distribution matrices, we divert the attention of the reader to Table 2 in \citet{sax2013temporal2}.
                                                                                                                                                                                                 
                                                                                                                                                                                                 \subsection{Extension to high-dimensional settings\label{sec:Extension to high-dimensional settings}}
                                                                                                                                                                                                 
                                                                                                                                                                                                 The shortcoming of \citet{chow1971best} becomes evident in data-rich environments, where the number of indicators $d\gg n$ surpass that of the time-stamps for the low-frequency data. Let us once again recall the GLS estimator \eqref{eq:GLSEstimator}. When $d<n$ and the columns of $\bm{X}_q^{\top}\bm{V}_q^{-1}\bm{X}_q$ are independent, the estimator is well-defined. However, when $d>n$, the matrix is rank-deficient - i.e., $\text{rank}(\bm{X}_q^{\top}\bm{V}_q^{-1}\bm{X}_q)\leq \min(n,d)$, the matrix $\bm{X}_q^{\top}\bm{V}_q^{-1}\bm{X}_q$ has linearly dependent columns, and thus is not invertible. In moderate dimensions, where $d\approx n$, $\bm{X}_q^{\top}\bm{V}_q^{-1}\bm{X}_q$ has eigenvalues close to zero, leading to high variance estimates of $\hat{\bm{\beta}}$.
                                                                                                                                                                                                 \citet{10.1111/rssa.12952} resolve this problem by adding a regularising penalty (e.g., $\ell_1$ regulariser) onto the GLS cost function \eqref{eq:CostFunction}:
                                                                                                                                                                                                   \begin{equation}
                                                                                                                                                                                                 \label{eq:LASSO}
                                                                                                                                                                                                 \hat{\bm{\beta}}(\lambda_n\mid\rho)=\arg\min_{\bm{\beta}\in\R^d}\left\{\Big\lVert \bm{V}_q^{-\frac{1}{2}}(\bm{y}_q-\bm{X}_q\bm{\beta})\Big\rVert_2^2+\lambda_n\lVert \bm{\beta}\rVert_1\right\}
                                                                                                                                                                                                 \end{equation}
                                                                                                                                                                                                 
                                                                                                                                                                                                 Unlike the GLS estimator \eqref{eq:GLSEstimator}, the regularised estimator corresponding to the cost function \eqref{eq:LASSO} is a function of $\lambda_n$ and the autoregressive parameter $\rho$. Henceforth, it is important to nominate the most suitable $\lambda_n$ and $\rho$ to correctly recover the parameters. In \eqref{eq:LASSO}, we denote the estimator as $\hat{\bm{\beta}}(\lambda_n\mid\rho)$ to highlight that the solution paths of the estimator for different values of $\lambda_n$, say, $\lambda_n^{(1)},\lambda_n^{(2)},\cdots,\lambda_n^{(k)}$ are generated for (i.e. conditional on) a fixed $\rho$. The solution paths are obtained using the LARS algorithm proposed \citet{efron2004least}, the benefits of which have been extensively discussed in \citet{10.1111/rssa.12952}.
                                                                                                                                                                                                 
                                                                                                                                                                                                 LASSO estimators inherently exhibit a small bias, such that $\lVert \hat{\bm{\beta}}\rVert_2^2\leq \lVert \bm{\beta}^*\rVert_2^2$, where $\bm{\beta}^*$ denotes the true coefficient vector. To alleviate this issue, \citet{10.1111/rssa.12952} further follow \citet{belloni2013least}, by performing a refitting procedure using least squares re-estimation. The latter entails generating a new $n\times d^{(l)}$ sub-matrix $\bm{X'}_q$, where $d^{(l)}\leq d$ from the original $n\times d$ matrix $\bm{X}_q$, with $\bm{X'}_q$ corresponding to the columns of $\bm{X}_q$ supported by $\hat{\bm{\beta}}(\lambda_n^{(l)}\mid\rho)$, for solutions $l=1,\cdots,k$ obtained from the LARS algorithm\footnote{noting the LARS algorithm produces solutions evaluated at a series of $\{\lambda_l\}_{l=1}^k$ points.}. We then perform a usual least squares estimation on $(\bm{y}_q,\bm{X'}_q)$ to obtain debiased solution paths for each $\lambda_n^{(l)}$.    

Finally, \citet{10.1111/rssa.12952} choose the optimal estimate from $\hat{\beta}(\lambda_n^{1}\mid \rho),\cdots,\hat{\beta}(\lambda_n^{k}\mid \rho)$ using the Bayesian Information Criterion (BIC hereafter) proposed by \citet{schwarz1978estimating}. The motivation for nominating this statistic over resampling methods, such as cross-validation or bootstrapping techniques, stems from the small sample size in the low frequency observations. The optimal regularisation is chosen conditional on $\rho$ according to
\begin{equation}
\label{eq:BIC}
\hat{\lambda}_n(\rho)={\arg\min}_{\lambda_n(\rho)\in\{\lambda_n^{(1)}(\rho),\cdots,\lambda_n^{(k)}(\rho)\}}\left\{-2\mathcal{L}\left(\hat{\bm{\beta}}(\lambda_n\mid \rho),\hat{\sigma}^2\right)+\log(n)K_{\lambda_n(\rho)}\right\}
\end{equation}
where $K_{\lambda_n(\rho)}=\lvert \{r:(\hat{\beta}(\lambda_n\mid \rho)_r\neq 0)\}\rvert$ is the degrees of freedom and $\mathcal{L}(\hat{\bm{\beta}}(\lambda_n\mid \rho),\hat{\sigma}^2)$ is the log-likelihood function of the GLS regression \eqref{eq:GLSEstimator}, 

%
which in the presence of Gaussian errors, is given by:
\begin{equation}
\mathcal{L}(\hat{\bm{\beta}},\hat{\sigma}^2)=-\frac{n}{2}\log(2\pi)-\frac{n}{2}\log(\sigma^2)-\frac{1}{2}\log(\lvert \bm{S}\rvert)-\frac{1}{2\sigma^2}(\bm{y}_q-\bm{X}_q\bm{\beta})^\top(\bm{y}_q-\bm{X}_q\bm{\beta}),
\end{equation}
where $\lvert \bm{S}\rvert$ is the determinant of the Toeplitz matrix $\bm{S}$ depending on $\rho$, such that $\bm{V}_q=\sigma^2\bm{S}$.
\FloatBarrier
\section{The \textquotedblleft{}DisaggregateTS\textquotedblright{} package\label{sec:The DisaggregateTS package}}

In this Section, we showcase the main functions that has been included in the \textbf{DisaggregateTS} package. Following \citet{sax2016package}, we first introduce the main function of the package and it its features, and subsequently we will showcase other functions that allow the practitioner to conducting simulations and analyses.  

\subsection{Functions\label{sec:Functions}}

The main function of the package which performs the sparse temporal disaggregation method proposed by \citet{10.1111/rssa.12952} is \texttt{disaggregate()}. This function is of the following form:

\begin{Schunk}
\begin{Sinput}
> disaggregate(Y, X, aggMat, aggRatio, method, ...)
\end{Sinput}
\end{Schunk}
\noindent where the first argument of the function, \texttt{Y}, corresponds to the $n\times 1$ vector of low-frequency series $\bm{y}_q$ that we wish to disaggregate, and the second argument, \texttt{X}, is the $p\times d$ matrix of high-frequency indicator series $\bm{X}_m$. In the event that there is no input \texttt{X}, the disaggregation matrix $\bm{X}_m$ is replaced with an $n\times 1$ vector of ones. 

The argument \texttt{aggMat} coincides with the aggregation matrix $\bm{C}$ in \eqref{eq: flowData}, and it has been set to \texttt{"sum"} by default, rendering it suitable for flow data. Alternative options include \texttt{"first"}, \texttt{"last"} and \texttt{"average"}. The aggregation (distribution) matrices that are utilised in this function are summarized in table 2 of \citet{sax2016package}.
The argument \texttt{aggRatio} has been set to \texttt{4} by default, which represents the ratio of annual to quarterly data. In general, this argument should be set to the ratio of the high-frequency to low-frequency series. For instance, in the examples considered in the preceding Sections, we had considered quarterly data as the low-frequency series, and monthly data as its high-frequency counterpart. Thus, in this setting \texttt{aggRatio = 3}. At first glance, the presence of the \texttt{aggRatio} argument may seem redundant. However, if $n\geq n_l\times \text{aggRatio}$, then extrapolation is done up to $n$.  

Finally, the argument \texttt{method} refers to the method of disaggregation under consideration. This argument has been set to \texttt{"Chow-lin"} method by default, which is the classical regression-based disaggregation technique introduced in Section \ref{sec:Classical regression-based techniques}. Other classical low-dimensional options include \texttt{"Denton"}, \texttt{"Denton-Cholette"}, \texttt{"Fernandez"}, and \texttt{"litterman"}, where these techniques have been extensively discussed in \citet{dagum2006benchmarking} and \citet{sax2013temporal2}. The main contribution of this package stem from the \texttt{"spTD"} and \texttt{"adaptive-spTD"} options pertaining to sparse temporal disaggregation and adaptive sparse temporal disaggregation, which are \citet{10.1111/rssa.12952}'s high-dimensional extension of the regression-based techniques proposed by \citet{chow1971best}. In a high-dimensional regression, the adaptive LASSO is relevant when, for instance, the columns of the design matrix $\bm{X}$ exhibit multicollinearity, and the \textit{Irrepresentability Condition} (IC hereafter) is violated (see \citet{zou2006adaptive} for details). In such settings, the regularization parameter $\lambda$ does not satisfy the oracle property, which can lead to inconsistent variable selection. The adaptive counterpart of the the regularized GLS cost function \eqref{eq:LASSO}, can be expressed as follows:
                                                                                                                                                                                                     \begin{equation}
                                                                                                                                                                                                   \label{eq:AdaptiveLASSO}
                                                                                                                                                                                                   \hat{\bm{\beta}}(\lambda_n\mid\rho)=\arg\min_{\bm{\beta}\in\R^d}\left\{\Big\lVert \bm{V}_q^{-\frac{1}{2}}(\bm{y}_q-\bm{X}_q\bm{\beta})\Big\rVert_2^2+\lambda_n\sum_{j=1}^{d}\frac{\lvert \beta_j\rvert}{\lvert \hat{\beta}_{\text{init},j}\rvert}\right\}
                                                                                                                                                                                                   \end{equation}
                                                                                                                                                                                                   where $\hat{\beta}_{\text{init},j}$ is an initial estimator, predicated on $\hat{\bm{\beta}}(\hat{\rho})$ from the regularized (LASSO) temporal disaggregation. See \citet{10.1111/rssa.12952}, for the details of the proposed methodology, and \citet{zou2006adaptive} and \citet{van2011adaptive} to yield variable selection consistency using the OLS estimator and LASSO as $\hat{\beta}_{\text{init},j}$  when the IC condition is violated.
                                                                                                                                                                                                   
                                                                                                                                                                                                   The second main function of the \textquotedblleft DisaggregateTS\textquotedblright{} package is \texttt{TempDisaggDGP()}, which generates synthetic data that can be used for conducting simulations using the \texttt{disaggregate()} function. The main arguments of this function are as follows:
                                                                                                                                                                                                     
\begin{Schunk}
\begin{Sinput}
> TempDisaggDGP(n_l, n, aggRatio, p, beta, sparsity, method, aggMat, rho, ...)
\end{Sinput}
\end{Schunk}
                                                                                                                                                                                                     \noindent where the first argument corresponds to the size of low-frequency series and \texttt{n} to that of the high-frequency series. Moreover, \texttt{aggRatio} and \texttt{aggMat} are defined as before, in turn representing the ratio of the high-frequency to low-frequency series, as well as the aggregation matrix \eqref{eq: flowData}. A minor difference in the DGP function is that if $n\geq n_l\times \text{aggRatio}$, then the last $n-\text{aggRatio}\times n_l$ columns of the aggregation matrix are set to zero, such that $Y$ is observed only up to $n_l$. Argument \texttt{p} sets the dimensionality of high-frequency series (set to $1$ by default), \texttt{beta} which has been set to $1$ by default is the positive and negative elements of the coefficient vector, \texttt{sparsity} is the sparsity percentage of the coefficient vector, and 
                                                                                                                                                                                                   \texttt{rho} is the autocorrelation of the error terms, which has been set to $0$ by default. Finally, the \texttt{method} argument determines the data generating process of the error terms, corresponding to methods discussed earlier in this Section. 
                                                                                                                                                                                                   
                                                                                                                                                                                                     
                                                                                                                                                                                                     A number of optional arguments included in the function determine the mean vector and the standard deviation of the design matrix, as well as options such a setting seed for running the simulations, where the design matrix and the coefficient vectors are fixed.  
                                                                                                                                                                                                   
                                                                                                                                                                                                   In what follows, we showcase a simple example of the function and its respective outputs:
                                                                                                                                                                                                     
\begin{Schunk}
\begin{Sinput}
> # Load the DisaggregateTS library
> library(DisaggregateTS)
> # Generate low-frequency quarterly series and its high-frequency monthly counterpart
> SynthethicData <- TempDisaggDGP(n_l = 2, n = 6, aggRatio = 3,
+ p = 6, beta = 0.5, sparsity = 0.5, method  = 'Chow-Lin', rho = 0.5)
\end{Sinput}
\end{Schunk}
                                                                                                                                                                                                     In the example above, we generate low-frequency series $\bm{y}_q\in \mathbb{R}^2$ corresponding to two quarters, and consequently, its high-frequency monthly counterpart $\bm{y}_m \in \mathbb{R}^{6}$. It is further assumed that the data is generated using six monthly indicators - i.e., $\bm{X}^{6\times 6}_m$, with a coefficient vector $\bm{\beta}\in \mathbb{R}^6$, where $\beta_j \in \{-0.5, 0, +0.5\}$. Since, the sparsity argument is set to $0.5$, only half of $\bm{\beta}$'s elements are non-zero. Finally, the error vector $\bm{u}_m$ is assumed to follow the structure AR($1$) structure of \citet{chow1971best}, with an autocorrelation parameter of $\rho=0.5$. 

\subsection{Simulations}\label{sec:Simulations}
In this Section, we show a simulation exercise to demonstrate the implementation of the temporal disaggregation method via the \textbf{DisaggregateTS} package.

\subsubsection*{Classic setting}

We start by simulating the dependent variable $Y\in \mathbb{R}^{17}$ and the set of high-frequency exogenous variables $X \in \mathbb{R}^{68\times 5}$ by using the command:

\begin{Schunk}
\begin{Sinput}
> # Load the DisaggregateTS library
>  library(DisaggregateTS)
>  # Generate low-frequency yearly series and its high-frequency quarterly counterpart
>  n_l = 17 # The number of low-frequency data points
> - annual
>  n = 68 # The number of high-frequency (quarterly) data points.
>  p_sim = 5 # The number of the high-frequency exogenous variables.
>  rho_sim = 0.8 # autocorrelation parameter
>  Sim_data <- TempDisaggDGP(n_l, n, aggRatio = 4, p = p_sim, rho = rho_sim)
>  Y_sim <- matrix(Sim_data$Y_Gen) # Extract the simulated dependent low-frequency variable
>  X_sim <- Sim_data$X_Gen # Extract the simulated exogenous high-frequency variables
\end{Sinput}
\end{Schunk}

In this example, we are generating a set of low-frequency data, i.e. 17 annual datapoints and a set of high-frequency (quarterly) exogenous variables that we want to use to infer the high-frequency counterpart of the low-frequency data. We now want to temporally disaggregate the low-frequency time series by using the information encapsulated in the high-frequency time series. In this case, since the number of time observations is larger than the number of exogenous variables, we can use standard methodologies to estimate the temporal disaggregation model. To do so, we use the \texttt{disaggregate} function setting \texttt{method="Chow-Lin"}. The code is as follows:

\begin{Schunk}
\begin{Sinput}
> C_sparse_SIM <- disaggregate(Y_sim, X_sim, aggMat = "sum", 
+ aggRatio = 4, method = "Chow-Lin")
> C_sparse_SIM$beta_Est
> Y_HF_SIM = C_sparse_SIM$y_Est[ ,1] # Extract the temporal disaggregated 
> # dependent variable estimated through the function disaggregate()
\end{Sinput}
\end{Schunk}

\begin{figure}[h]
    \centering
    \includegraphics[width=\textwidth]{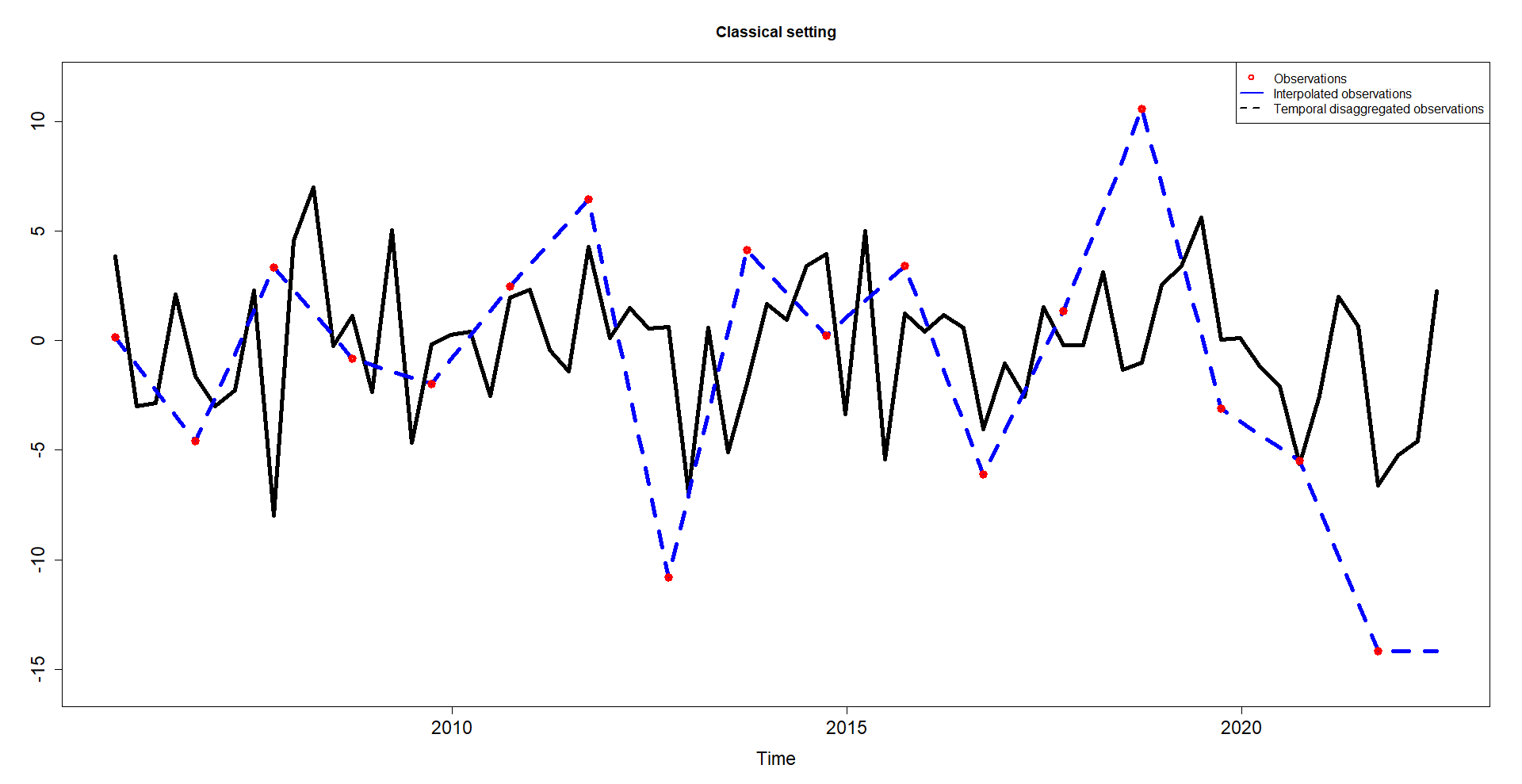}
    \caption{Temporal disaggregated and interpolated observations for the estimation under the classical setting. The plot is built using the snipped code provided in this subsection. As we used the setting \texttt{aggMat = "sum"}, the sum of every four disaggregated observations correspond to an actual observation.}
    \label{fig_res_sim_cs}
\end{figure}

We show in Figure \ref{fig_res_sim_cs} the results, where we depict the original (low-frequency) time series together with the high-frequency counterpart computed via standard interpolation and estimated through the Chow-Lin temporal disaggregation method.   

\subsubsection*{High-dimensional setting}
We now repeat the simulation experiment in a high-dimensional setting, where the number of temporal observations is lower than the number of exogenous variables. In this case, standard methods like Chow-Lin cannot be applied.
To do so, we simulate the dependent variable $Y\in \mathbb{R}^{17}$ as before, but now the set of high-frequency exogenous variables is of dimension $X \in \mathbb{R}^{68\times 100}$. Similarly, as before, we can use the following command:

\begin{Schunk}
\begin{Sinput}
>  # Load the DisaggregateTS library
>  library(DisaggregateTS)
>  # Generate low-frequency yearly series and its high-frequency quarterly counterpart
>  n_l = 17 # The number of low-frequency data points - annual 
>  n=68 # The number of high-frequency data points 0 quarterly
>  p_sim = 100 # The number of the high-frequency exogenous variables.
>  rho_sim = 0.8 # autocorrelation parameter
>  Sim_data <- TempDisaggDGP(n_l, n, aggRatio = 4,p = p_sim, rho = rho_sim)
>  Y_sim <- matrix(Sim_data$Y_Gen) #Extract the simulated dependent 
>  # (low-frequency) variable
>  X_sim <- Sim_data$X_Gen #Extract the simulated exogenous variables - high-frequency
\end{Sinput}
\end{Schunk}
In this case, we cannot use a standard technique, and we need to estimate a sparse model to overcome the curse of dimensionality. The \texttt{disaggregate} function can handle the high-dimensional setting by choosing the method to be \texttt{spTD} or \texttt{adaptive-spTD}. In the following example, we use the latter:

\begin{Schunk}
\begin{Sinput}
>  C_sparse_SIM = disaggregate(Y_sim, X_sim, aggMat = "sum", 
+  aggRatio = 4, method = "adaptive-spTD")
>  C_sparse_SIM$beta_Est
>  Y_HF_SIM = C_sparse_SIM$y_Est[ ,1] # Extract the temporal disaggregated 
> # dependent variable estimated through the function disaggregate()
\end{Sinput}
\end{Schunk}

Figure \ref{fig_res_sim_hds} below shows the results.

\begin{figure}[h]
    \centering
    \includegraphics[width=\textwidth]{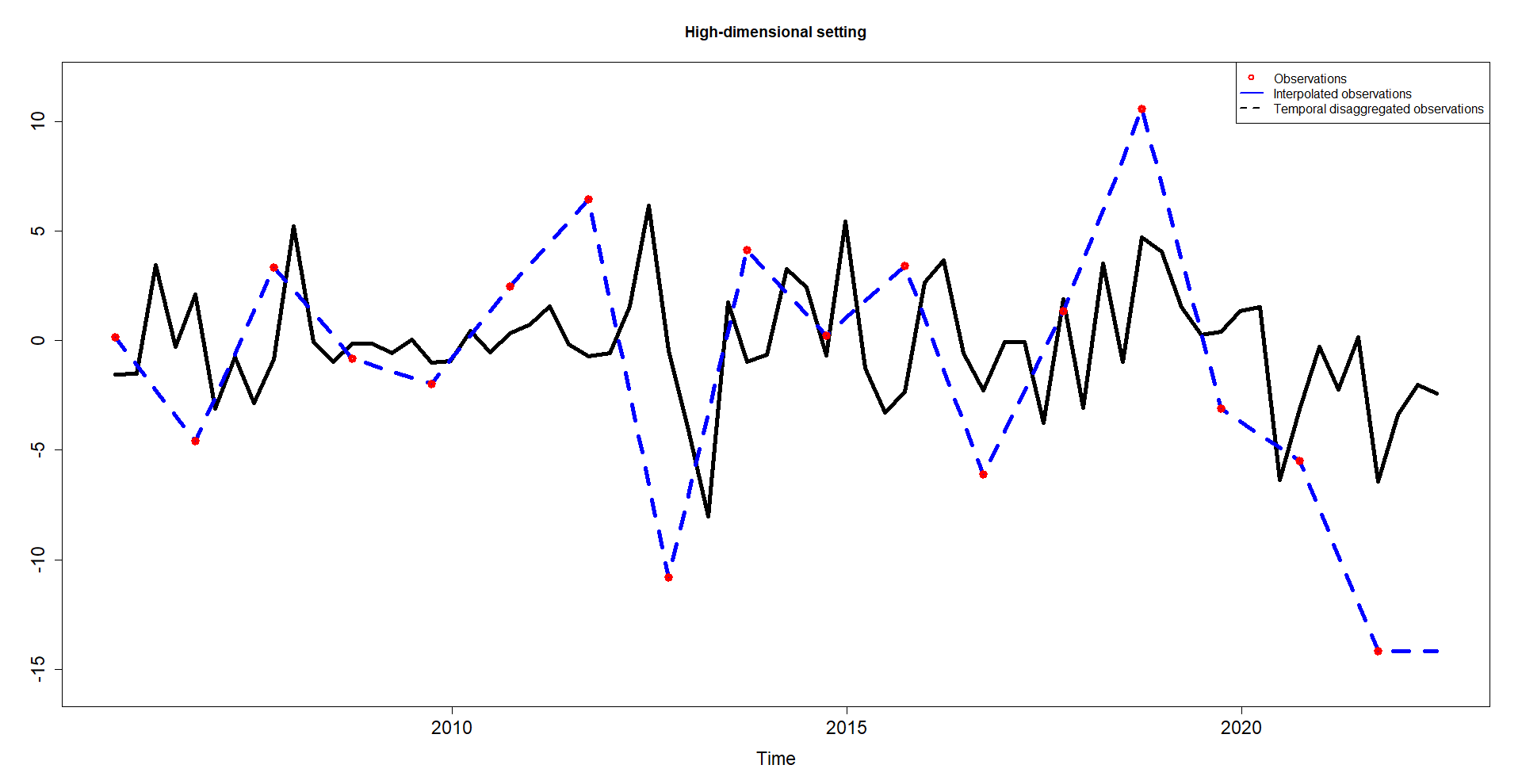}
    \caption{Temporal disaggregated and interpolated observations for the estimation under the high-dimensional setting. The plot is generated using the code snippet provided in this subsection. Specifically, with the setting \texttt{aggMat = "sum"}, we aggregate the data by summing every four consecutive observations. Each of these sums represents a single actual observation.}
    \label{fig_res_sim_hds}
\end{figure}

As we can see from both Figures \ref{fig_res_sim_cs} and \ref{fig_res_sim_hds}, the standard interpolation cannot reproduce the fluctuations of the data, making the result overly smooth. The temporal disaggregation method on the other hand, recreates reliable fluctuations.

\subsection{Empirical application: Carbon intensity}\label{sec:Applications}
In this Section, we show how temporal disaggregation can be used in a real-world problem.\\
The urgent need to address climate change has propelled the scientific community to explore innovative approaches to quantify and manage greenhouse gas (GHG) emissions. Carbon intensity, a crucial metric that measures the amount of carbon dioxide equivalent emitted per unit of economic activity (e.g. sales), plays a pivotal role in assessing the environmental sustainability of industries, countries, and global economies. Accurate and timely carbon accounting and the development of robust measurement frameworks are essential for effective emission reduction strategies and the pursuit of sustainable development goals. 
While carbon accounting frameworks offer valuable insights into emissions quantification, they are not without limitations. One of those limitations is the frequency with which this information is released, generally at an annual frequency, while most companies' economic indicators are made public on a quarterly basis. 
                                                                                                                                                                                                   This is a perfect example in which temporal disaggregation can be used to bridge the gap between data availability and prompt economic and financial analyses. In this application, the variable of interest is the GHG emissions for IBM between Q3 2005 and Q3 2021, at annual frequency, resulting in 17 datapoints, i.e. $Y\in \mathbb{R}^{17}$. For the high-frequency data, we used the balance sheet, income statement, and cash flow statement quarterly data between Q3 2005 and Q3 2021, resulting in 68 datapoints for the 128 variables. We remove variables that have a pairwise correlation higher than 0.99, resulting in a filtered dataset with 112 variables, i.e. $X \in \mathbb{R}^{68\times 112}$.
                                                                                                                                                                                                   
\begin{figure}[h]
    \centering
    \includegraphics[width=\textwidth]{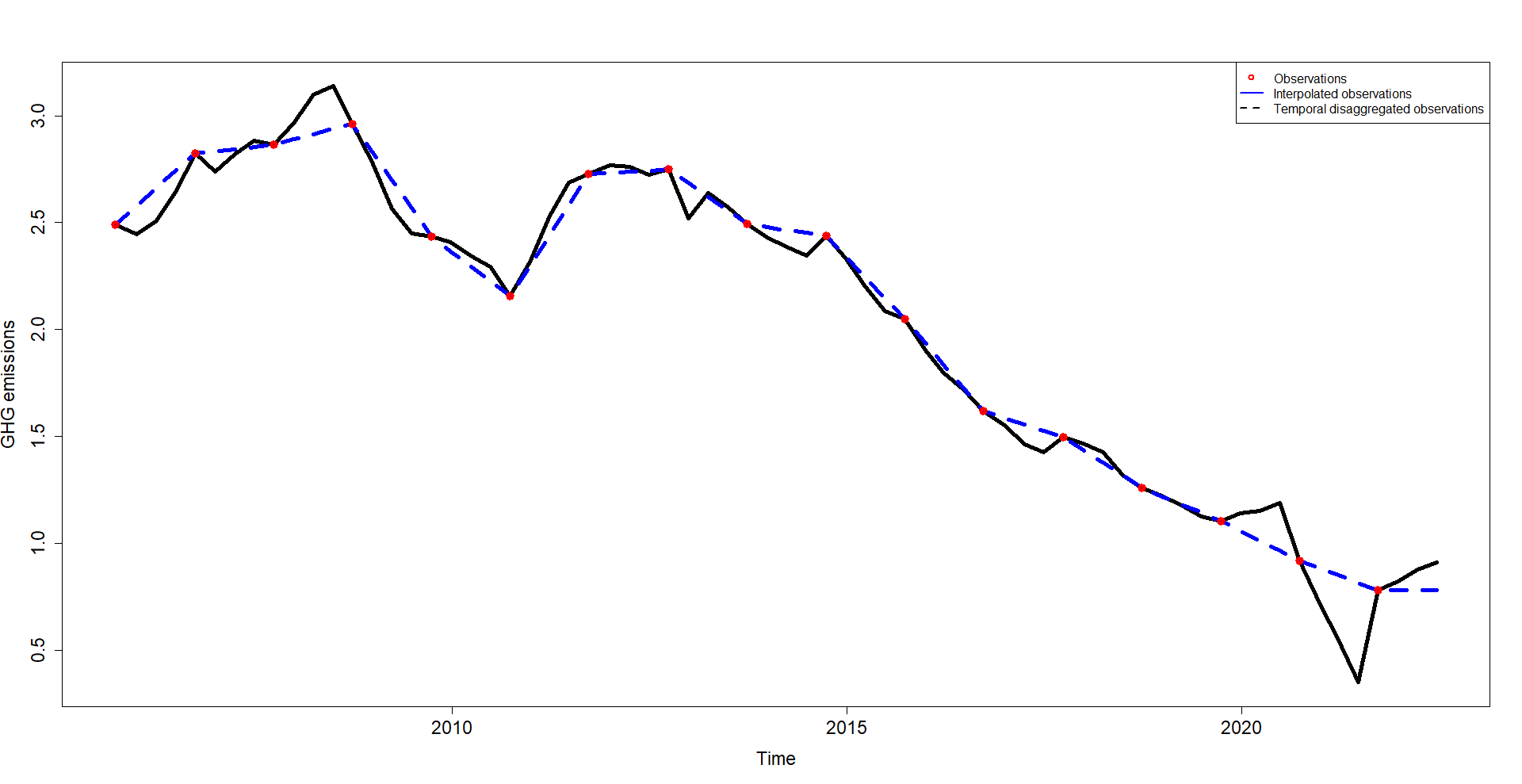}
    \caption{Temporal disaggregated and interpolated GHG emissions observations. In this example, we used the setting \texttt{aggMat = "first"}, so the first disaggregated is anchored correspond to the actual first observation.}
    \label{fig_res}
\end{figure}

                                                                                                                                                                                                   In this example, we employed the adaptive LASSO method, resulting in only two non-zero coefficients, which are the 12 months trailing sales and the total company capital. Trailing sales and total company capital are indeed relevant predictors of emissions because they reflect a company's economic activity, operational intensity, and commitment to sustainability. We show the results in Figure \ref{fig_res} alongside a linear interpolation method.

As it is possible to observe from the plot, the interpolated data do not fluctuate as we would expect from real GHG emissions, as the method is not conditional on the variability of the high-frequency variables. In this respect, the temporal disaggregated observations show a remarkably truthful dynamics. This result can then be used to compute the GHG intensity, computing the ratio between GHG emissions and the sales for the corresponding quarter.


\FloatBarrier
\section{Summary\label{sec:Summary}}
In this paper, we demonstrated how the \textbf{DisaggregateTS} package can be used and what are its potential use in climate finance and economics. The data-generating processes encoded within the model allow for efficient synthetic evaluation of disaggregation procedures.



\bibliographystyle{apalike}
\bibliography{main}

\begin{thebibliography}{}

\bibitem[Belloni and Chernozhukov, 2013]{belloni2013least}
Belloni, A. and Chernozhukov, V. (2013).
\newblock Least squares after model selection in high-dimensional sparse
  models.

\bibitem[Bournay and Laroque, 1979]{bournay1979reflexions}
Bournay, J. and Laroque, G. (1979).
\newblock R{\'e}flexions sur la m{\'e}thode d'elaboration des comptes
  trimestriels.
\newblock In {\em Annales de l'INSEE}, pages 3--30. JSTOR.

\bibitem[Chow and Lin, 1971]{chow1971best}
Chow, G.~C. and Lin, A.-l. (1971).
\newblock Best linear unbiased interpolation, distribution, and extrapolation
  of time series by related series.
\newblock {\em The review of Economics and Statistics}, pages 372--375.

\bibitem[Chow and Lin, 1976]{chow1976best}
Chow, G.~C. and Lin, A.-L. (1976).
\newblock Best linear unbiased estimation of missing observations in an
  economic time series.
\newblock {\em Journal of the American Statistical Association},
  71(355):719--721.

\bibitem[Dagum and Cholette, 2006]{dagum2006benchmarking}
Dagum, E.~B. and Cholette, P.~A. (2006).
\newblock Benchmarking, temporal distribution, and reconciliation methods for
  time series.

\bibitem[Denton, 1971]{denton1971adjustment}
Denton, F.~T. (1971).
\newblock Adjustment of monthly or quarterly series to annual totals: an
  approach based on quadratic minimization.
\newblock {\em Journal of the american statistical association},
  66(333):99--102.

\bibitem[Efron et~al., 2004]{efron2004least}
Efron, B., Hastie, T., Johnstone, I., and Tibshirani, R. (2004).
\newblock Least angle regression.

\bibitem[Fernandez, 1981]{fernandez1981methodological}
Fernandez, R.~B. (1981).
\newblock A methodological note on the estimation of time series.
\newblock {\em The Review of Economics and Statistics}, 63(3):471--476.

\bibitem[Fuleky, 2019]{fuleky2019macroeconomic}
Fuleky, P. (2019).
\newblock {\em Macroeconomic forecasting in the era of big data: Theory and
  practice}, volume~52.
\newblock Springer.

\bibitem[Litterman, 1983]{litterman1983random}
Litterman, R.~B. (1983).
\newblock A random walk, markov model for the distribution of time series.
\newblock {\em Journal of Business \& Economic Statistics}, 1(2):169--173.

\bibitem[Mosley et~al., 2022]{10.1111/rssa.12952}
Mosley, L., Eckley, I.~A., and Gibberd, A. (2022).
\newblock {Sparse Temporal Disaggregation}.
\newblock {\em Journal of the Royal Statistical Society Series A: Statistics in
  Society}, 185(4):2203--2233.

\bibitem[Mosley and {S. Nobari}, 2022]{}
Mosley, L. and {S. Nobari}, K. (2022).
\newblock {\em DisaggregateTS: High-Dimensional Temporal Disaggregation}.
\newblock R package version 2.0.0.

\bibitem[Quilis, 2018]{quilis2018temporal}
Quilis, E.~M. (2018).
\newblock Temporal disaggregation of economic time series: The view from the
  trenches.
\newblock {\em Statistica Neerlandica}, 72(4):447--470.

\bibitem[Sax and Steiner, 2013]{sax2013temporal2}
Sax, C. and Steiner, P. (2013).
\newblock Temporal disaggregation of time series.

\bibitem[Sax et~al., 2016]{sax2016package}
Sax, C., Steiner, P., Di~Fonzo, T., and Sax, M.~C. (2016).
\newblock Package ‘tempdisagg’.

\bibitem[Schwarz, 1978]{schwarz1978estimating}
Schwarz, G. (1978).
\newblock Estimating the dimension of a model the annals of statistics 6 (2),
  461--464.
\newblock {\em URL: http://dx. doi. org/10.1214/aos/1176344136}.

\bibitem[Tibshirani, 1996]{tibshirani1996regression}
Tibshirani, R. (1996).
\newblock Regression shrinkage and selection via the lasso.
\newblock {\em Journal of the Royal Statistical Society: Series B
  (Methodological)}, 58(1):267--288.

\bibitem[Van~de Geer et~al., 2011]{van2011adaptive}
Van~de Geer, S., B{\"u}hlmann, P., and Zhou, S. (2011).
\newblock The adaptive and the thresholded lasso for potentially misspecified
  models (and a lower bound for the lasso).

\bibitem[Wainwright, 2019]{wainwright2019high}
Wainwright, M.~J. (2019).
\newblock {\em High-dimensional statistics: A non-asymptotic viewpoint},
  volume~48.
\newblock Cambridge university press.

\bibitem[Zou, 2006]{zou2006adaptive}
Zou, H. (2006).
\newblock The adaptive lasso and its oracle properties.
\newblock {\em Journal of the American statistical association},
  101(476):1418--1429.

\end{thebibliography}
\nocite{*}

\end{document}